# Explosive Chrysopoeia

Jan Maurycy Uszko[1], Stephen J. Eichhorn[1], Avinash J. Patil[2], Simon R. Hall[2*]

[1]The Bristol Composites Institute (BCI), School of Civil, Aerospace and Design Engineering, University Walk, University of Bristol, Bristol, BS8 1TR, UK

[2] School of Chemistry, University of Bristol, Bristol BS8 1TS, United Kingdom

*Corresponding author. Email: simon.hall@bristol.ac.uk

## Abstract

**Fulminating gold, the first high-explosive compound to be discovered, disintegrates in a mysterious cloud of purple smoke, the nature of which has been speculated upon since its discovery in 1585. In this work, we show that the colour of the smoke is due to the presence of gold nanoparticles.**

## Main

The alchemist's fascination with the transmutation of base metals into gold (chrysopoeia) led to the discovery of the first high-explosive compound, fulminating gold.[1] Its first synthesis was described in 1585 by Sebalt Schwärtzer, requiring four to five days to complete.[2] Since then, the process has been studied and improved to the point where fulminating gold can now be synthesized in minutes by simply mixing gold (III) compounds with ammonia.[1] The ease at which this material can be synthesized has stimulated research into it, leading to reviews of the current state of research into fulminating gold being periodically published by scientific journals from "Über die Stickstoffverbindungen des Goldes" in 1915 to "Fulminating Gold and Silver" in 2019.[1,3,4] Besides academic interest, the other motivator for research on fulminating gold has been safety. The danger of accidentally creating highly explosive, touch-sensitive side products resulted in warnings on the use of gold (III) compounds being printed in journals, editorial letters, chemical textbooks as far back as 1851, and safety manuals like the popular "Bretherick's Handbook of Reactive Chemical Hazards".[1,5-7] The interest in fulminating gold has even spread to digital media, with a video showing the synthesis of fulminating gold by Thomas De Prinse on his YouTube channel "Explosions&Fire" reaching almost 1 million views since it was first posted in 2020.[8] In this video, De Prinse repeats a supposition that is often stated in relation to the source of the unusual red or purple colouration of the smoke created in the detonation of fulminating gold, that it is due to the presence of gold nanoparticles. There is circumstantial evidence that the smoke consists of gold nanoparticles, as it has been used to coat objects in a purple/crimson patina as described in "Opera Chymica" by Glauber (*4,9*),[4,9] much in the same way that solutions of gold nanoparticles can be used to coat substrates with purple/red layers.[10] To date, there has however been no experimental verification of this. Here, we show for the first time that the explosion of fulminating gold creates gold nanoparticles, ranging in size from 10 to 300 nm. Furthermore, given the extreme rapidity of their creation, they are more isotropic than nanoparticles created by conventional methods.

A typical synthesis was as follows: chloroauric acid (20 mg, Sigma Aldrich) was dissolved in 1 ml of deionized water in a polypropylene tube. To this solution, ammonium hydroxide (28 - 30 w/w % Fisher Scientific) was added dropwise until an orange precipitate formed. The suspension was divided into four aliquots and each placed on separate aluminium foils to dry overnight in air at room temperature. After drying, samples of approximately 5 mg were detonated by the application of heat to the aluminium foil, whilst carbon-coated TEM grids (copper 200 mesh, Agar Scientific) were held above the foil to catch the resultant purple cloud.

Figure 1 shows a high-resolution TEM image of a single nanoparticle with visible lattice fringes, that have a spacing of 0.24 nm, consistent with the (111) crystal planes of Au. The selected area electron diffraction pattern illustrated in Fig. 2 and Table 1 confirms that they are indeed Au(0), as per the Joint Committee on Powder Diffraction Standards (JCPDS) card no. 04-0784.

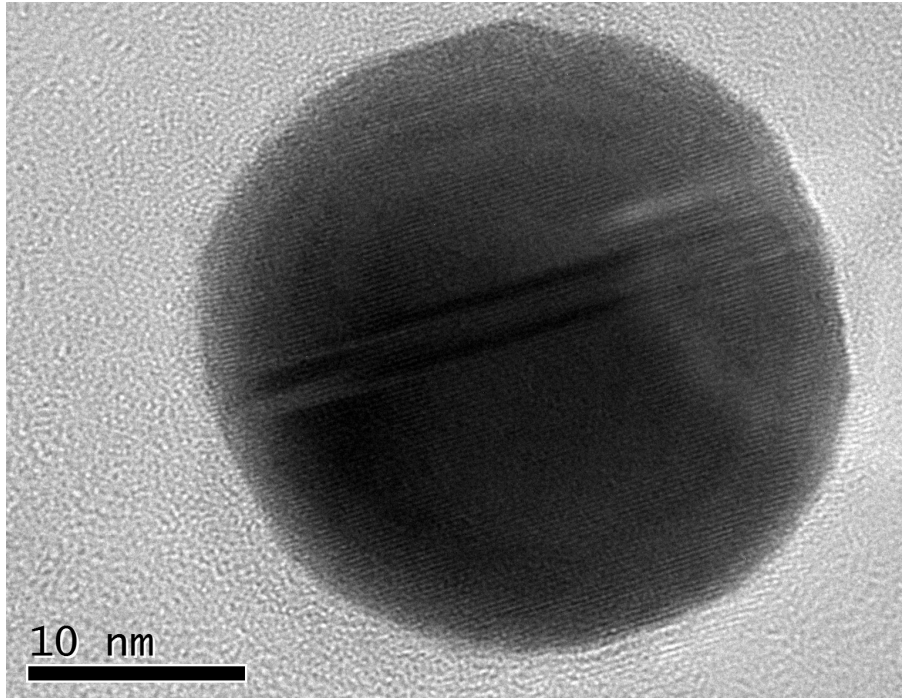

**Figure 1.** TEM image of a nanoparticle, from detonated fulminating gold, with visible lattice fringes

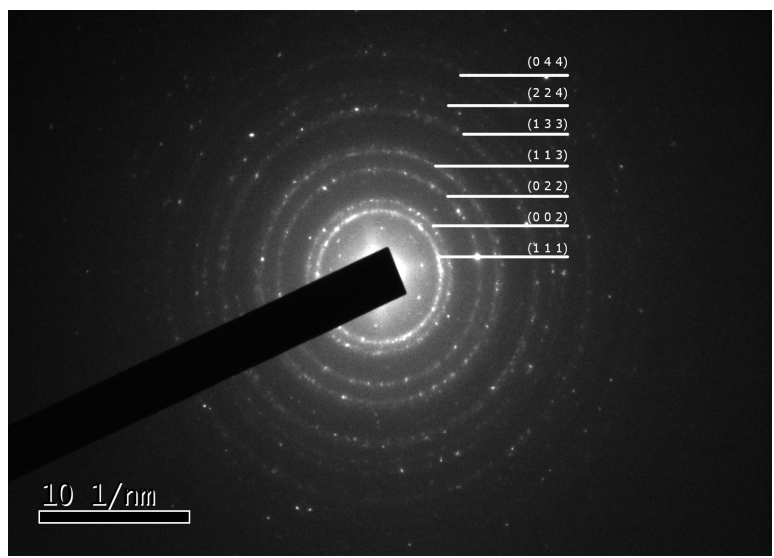

**Figure 2.** Selected Area Electron Diffraction ring pattern from gold nanoparticles. The rings are indexed to Au as per the Joint Committee on Powder Diffraction Standards (JCPDS) card no. 04-0784.

**Table 1.** Miller indices for electron diffraction rings with the theoretical (theor.) and measured radii and *d*-spacings of the rings.

| Ring identification (JCPDS #04-0784) | | | | |
|---|---|---|---|---|
| Plane | Radius [1/nm] | | *d*-spacing [nm] | |
| | theor. | measured | theor. | measured |
| (1 1 1) | 4.245 | 4.072 | 0.236 | 0.246 |
| (0 0 2) | 4.902 | 4.668 | 0.204 | 0.214 |
| (0 2 2) | 6.932 | 6.705 | 0.144 | 0.149 |
| (1 1 3) | 8.129 | 7.847 | 0.123 | 0.127 |
| (1 3 3) | 10.684 | 10.380 | 0.094 | 0.096 |
| (2 2 4) | 12.007 | 11.522 | 0.083 | 0.087 |
| (2 2 4) | 12.007 | 12.217 | 0.083 | 0.082 |
| (0 4 4) | 13.865 | 13.956 | 0.072 | 0.072 |

TEM images of grids that were exposed to the purple cloud showed clusters of spherical nanoparticles exhibiting a wide size distribution from 30 nm to over 300 nm, with an average particle diameter of 40 nm [σ = 44] illustrated in Figs. 3 and 4. The broad particle size distribution is indicative of the extreme rapidity of synthesis, with no possibility of achieving a lower polydispersity via the usual mechanisms of Ostwald ripening or through ligand passivation.[11,12] The absence of well-defined facets in the nanoparticles is intriguing and indicates the accelerated synthesis. The formation of the usual faceted or even triangular morphology of Au nanoparticles is effectively prevented through their creation by detonation. In this way, larger gold nanoparticles can be created with a sphericity more commonly seen in the early stages of formation when the nanoparticles are small.[12]

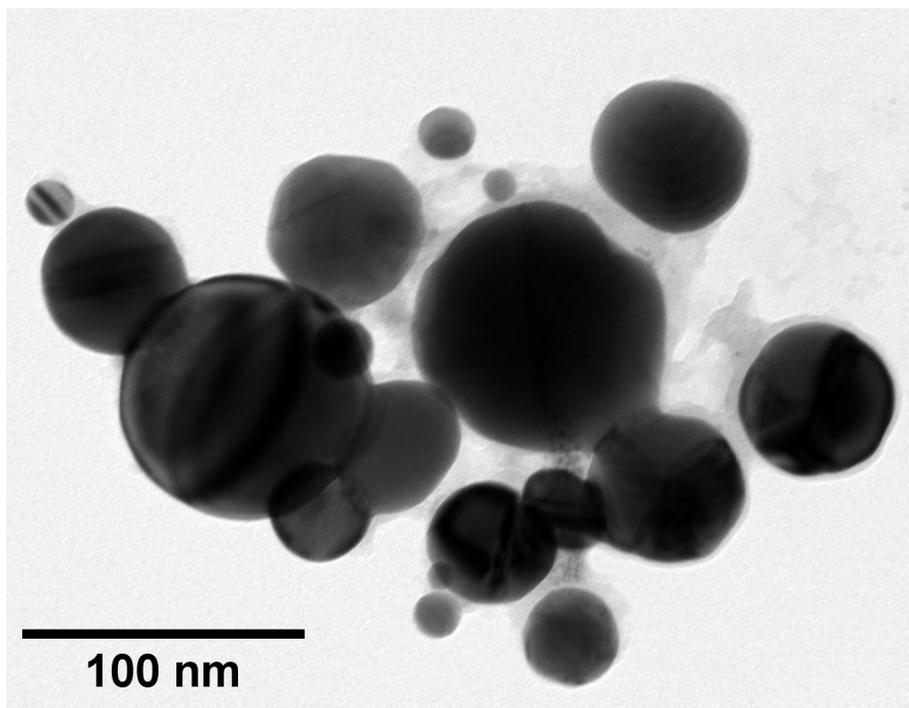

**Figure 3.** TEM image of a cluster of gold nanoparticles captured from detonated fulminating gold. The image demonstrates visually the broad dispersion of particle sizes.

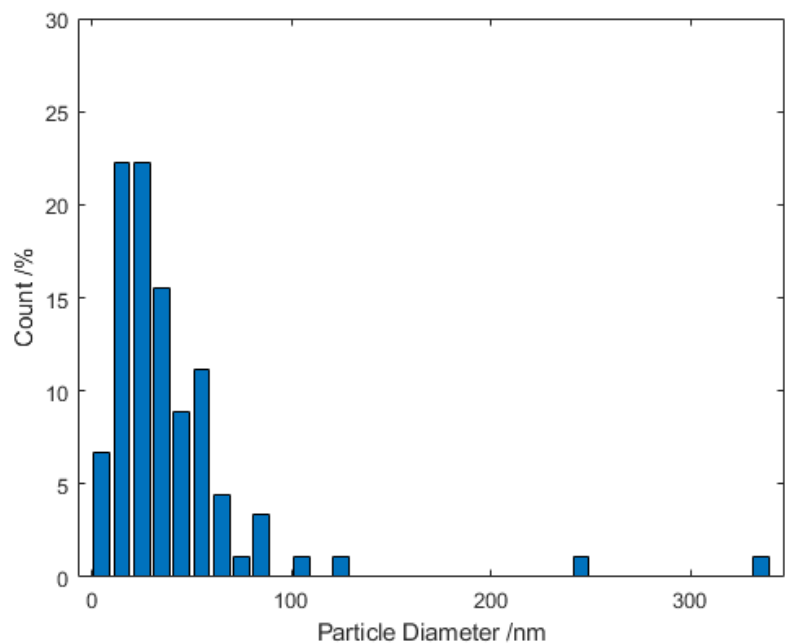

**Figure 4.** Particle size distribution of gold nanoparticles from the detonation of fulminating gold.

This work is proof of the long-supposed nature of the cloud produced on the detonation of fulminating gold, but also potentially opens the door to fast solvent- and capping agent-free syntheses of metal nanoparticles.

**Acknowledgments:**

EM studies were carried out in the Chemical Imaging Facility at the University of Bristol, with equipment funded by EPSRC under Grant "Atoms to Applications" Grant ref. (EP/K035746/1).

The authors would like to thank Carla Forster for her assistance with the translation of original sources.

**Funding:**

This work was funded by the EPSRC Centre for Doctoral Training in Composites Science, Engineering and Manufacturing (EP/S021728/1)

**Author contributions:**

Conceptualization: JMU, SRH

Methodology: JMU

Investigation: JMU

Visualization: JMU, SRH

Funding acquisition: SJE

Project administration: SRH, SJE, AJP

Supervision: SRH, SJE AJP

Writing – original draft: JMU, SRH

Writing – review & editing: JMU, SRH, SJE, AJP

**Competing interests:** Authors declare that they have no competing interests.

**Data and materials availability:** All data are available in the main text